\documentclass[showpacs,aps,prb,reprint,twocolumn]{revtex4-1}

\usepackage{commath}
\usepackage{amsmath,bm}
\usepackage{graphicx}
\usepackage{hyperref}
\usepackage{txfonts}

\draft 

\begin{document}


\title{On the electrical conductivity of metals with a rough surface}
\author{Hai-Yao Deng}
\email{haiyao.deng@gmail.com}
\affiliation{School of Physics and Astronomy, Cardiff University, 5 The Parade, Cardiff CF24 3AA, Wales, United Kingdom}
\begin{abstract}
We discuss surface roughness effects on the conduction of electrons in metals using both the quantal Kubo-Greenwood formalism and the semi-classical Fuchs-Sondheimer method. The main purpose here is to compare these methods and clarify a few subtle conceptual issues. One of such issues is concerned with the conditions under which the broken translation symmetry along a rough surface may be restored. This symmetry has often been presumed in existing work but not always with proper justifications. Another one relates to the physical meaning of a phenomenological parameter (denoted by $p$) intuitively introduced in the semi-classical theory. This parameter, called the specularity parameter or sometimes the \textit{Fuchs} parameter, plays an important role in the experimental studies of surface roughness but has so far lacked a rigorous microscopic definition. The third issue arises as to the domain of validity for the electrical conductivity obtained in those methods. A misplacement of the domain may have resulted in erroneous analysis of surface effects in a variety of electrodynamic phenomena including surface plasma waves.  
\end{abstract}
\maketitle

\section{introduction}
\label{sec:1}
Surface roughness can strongly affect electronic conduction in metals~\cite{abrikosov,ziman}. Two general analytically amenable  approaches have been used for dealing with surface roughness effects. The first is based on the semi-classical Boltzmann transport equation, proposed by Fuchs~\cite{fuchs1938} for analyzing the electrical conductivity of metal films and employed in Sondheimer's work~\cite{sondheimer1948,sondheimer1952} on anomalous skin effect~\cite{kaganov1997}. While remaining a standard reference for sorting out experimental observations~\cite{kuan2004,gall2007,gall2009} and benchmarking \textit{ab initio} calculations on the conductivity of metallic films~\cite{gall2008,ke2009}, this Fuchs-Sondheimer method has also been employed in the study of giant magneto-resistance in layered structures~\cite{camley1989,barnas1990,levy1990}, surface energy absorption profile~\cite{flores1979}, surface plasmon oscillations~\cite{harris1972,garcia1977,deng2019,deng2017a,deng2017b,deng2017c}, dynamical responses~\cite{deng2018}, and graphene plasmonic structures~\cite{katsnelson2018}. 

The second approach builds upon the quantal Kubo-Greenwood formalism~\cite{kubo,greenwood}. This approach, though not widely utilized for handling surface roughness, is arguably exact within the regime of linear responses. It has also been used in the analysis of giant magneto-resistance observed in layered structures~\cite{zhang1992} and for accounting for quantum size effects of surface scattering in very thin films~\cite{tesanovic1986}. One should also mention some additional approaches that have appeared in the literature~\cite{chambers1950,chambers1952,camblong1992,camblong1993,camblong1995}. These, however, may be looked upon as certain limits of the general approaches~\cite{zhang1995}.

In the present work, we raise and clarify a few important conceptual issues that have not been paid sufficient attention in the literature on these approaches, in order to improve our current understanding of them. We are primarily concerned with three issues: one about the loss of translational symmetry, another about the physical meaning of the specularity parameter~\cite{fuchs1938}, and the third about the domain of validity of the electrical conductivity calculated by the approaches. They are described with more details in what follows.

\textit{Restoration of translational symmetry along a rough surface}. The presence of a rough surface not only breaks the translational symmetry along the normal of the surface, but also the symmetry along the surface. Nevertheless, in the Fuchs-Sondheimer method, the planar symmetry is explicitly presumed in their solutions to Boltzmann's transport equation. In the Kubo-Greenwood method, the view is split: some authors simply presume such symmetry without justifications~\cite{zhang1995} while other authors restore the the symmetry by an averaging procedure over the surface profile~\cite{tesanovic1986,zhang1992}. The latter is reasonable in the calculation of the conductivity to a uniform electric field but needs further justifications otherwise. In particular, it needs to be understood how the explicit preservation of the planar symmetry in the Fuchs-Sondheimer method is consistent with the Kubo-Greenwood method. 

\textit{Microscopic definition of the specularity parameter}. In the Fuchs-Sondheimer method, the planar translational symmetry is retained and the conductivity is uniquely determined up to a single parameter $p$, which is envisaged to characterize surface roughness effects. According to Fuchs and his followers, $p$ gives the fraction of electrons impinging on a surface gets specularly reflected back. While intuitively appealing, this specularity parameter has never been given a microscopic definition. However, such a definition is important, because the parameter is often the quantity that is extracted in experimental ~\cite{kuan2004,gall2007,gall2009} and \textit{ab initio} computational studies~\cite{gall2008,ke2009}.

\textit{Domain of validity}. One reason that the planar translational symmetry seems obvious in the Fuchs-Sondheimer method is because Boltzmann's equation used in this method does not include anything explicit of a surface. In particular, it does not include the potential that confines the electrons to the metal. The conductivity obtained by this method therefore cannot be valid for the surface region where the confining potential varies significantly. As far as we are concerned, few existing work based on this method has warded off this fallacy. Ignorance of this might have led to erroneous results that have permeated widely~\cite{deng2019}. 

The main purpose of this work is to address the aforementioned issues. Firstly, we clarify the circumstances for the restoration of translational symmetry along a rough surface from both the quantum mechanical and semi-classical point of view. Secondly, we provide two related microscopic definitions of the specularity parameter $p$, one based on the Kubo-Greenwood formalism while the other on a more complete Boltzmann's equation. Finally and most importantly, we demarcate the domain of validity for the conductivity obtained with the Fuchs-Sondheimer method, and prescribe a formula to extend its validity to the entire system. We also discuss some confusions in existing work in light of the results. 

In the next section, we define our system and recapitulate the Fuchs-Sondheimer method in order to see the issues in formal terms and introduce some notations. In Sec.~\ref{sec:3}, we calculate the conductivity by an extended Boltzmann's equation that explicitly includes surface effects (Sec.~\ref{sec:3.1}) and by the Kubo-Greenwood formula (Sec.~\ref{sec:3.2}). We discuss the results and summarize the paper in Sec.~\ref{sec:4}. 

\section{Review of the Fuchs-Sondheimer method}
\label{sec:2}
\begin{figure}
\begin{center}
\includegraphics[width=0.47\textwidth]{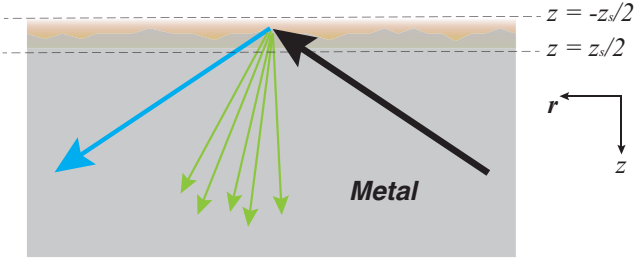}
\end{center}
\caption{Schematic of the system: a semi-infinite metal bounded by a rough surface. The surface layer has a thickness of $z_s$ and is characterized by a potential $V_s(\mathbf{x})$ that confines the electrons to the metal. The resulting force $\mathbf{F}_s(\mathbf{x}) = -\partial_\mathbf{x} V_s(\mathbf{x})$ is significant only in the surface region. A ``macroscopically flat" surface is defined as one for which the Fourier components, denoted by $V^{\mathbf{k}}_s(z)$ of $V_s(\mathbf{x})$ are negligible except for $k = \abs{\mathbf{k}}$ being zero or comparable to the inverse of a lattice constant, which is of the order of the electronic Fermi wavelength $k_F$. An electron wave incident on the surface is reflected either specularly (due to components with $k\approx0$) or diffusively (due to components with $k\sim k_F$). \label{fig:1}}
\end{figure}

The system studied throughout this paper is a semi-infinite metal shown in Fig.~\ref{fig:1}, modeled as an electron gas moving in a background of uniformly distributed positive charges (i.e. the jellium model). The electrons are confined to the metal by a potential $V_s(\mathbf{x})$, where $\mathbf{x} = (\mathbf{r},z)$ denotes the vector of a point in space with $\mathbf{r} = (x,y)$ being projection onto the surface. The confining potential varies significantly only within a narrow layer $z\in [-z_s/2,z_s/2]$, which defines the surface region with a thickness of $z_s>0$. In other words, the force $\mathbf{F}_s(\mathbf{x}) = - \partial_\mathbf{x}V_s(\mathbf{x})$ is negligible outside the surface region. The bulk of the metal then corresponds to $z \in (z_s/2, \infty)$ while the vacuum to $z \in (-\infty,-z_s/2)$. It is to be understood that $z_s$ is very small in comparison with a scale $\Lambda$ that is macroscopic in comparison with the lattice constant of the metal which we denote by $a$. Typically, $z_s$ is only a few lattice constants whereas $\Lambda$ exceeds a few nanometers, so that $z_s/\Lambda\ll1$. Semi-classical methods are supposed to work only on the scale of $\Lambda$.  

In the Fuchs-Sondhermier method, the surface serves as a geometric confining plane (i.e. $z_s\rightarrow0$) located at $z=0$ and the starting point is the following linearized Boltzmann equation within the relaxation-time approximation,
\begin{equation}
\left(\mathcal{D}_t + \mathbf{v}\cdot \partial_\mathbf{x}\right)g(\mathbf{x},\mathbf{v},t) = - \frac{e\mathbf{E}(\mathbf{x},t)}{m}\cdot \partial_{\mathbf{v}} f_0\left(\varepsilon(\mathbf{v})\right), \label{1}
\end{equation} 
where $\mathcal{D}_t = \partial_t + 1/\tau$ with $\tau$ being the relaxation time, $\mathbf{E}$ is the electric field felt by the electrons~\cite{note1}, $\varepsilon(\mathbf{v}) = m\mathbf{v}^2/2$ is the kinetic energy of an electron with $m$, $e$ and $\mathbf{v} = (\mathbf{v}_\parallel,v_z)$ being its mass, charge and velocity, respectively, and $g$ and $f_0$ are the non-equilibrium and equilibrium distribution (i.e. the Fermi-Dirac function) of the electrons, respectively. For later use, we also introduce $\mathbf{v}_\parallel = (v_x,v_y)$ as the planar projection of $\mathbf{v}$. 

Note that in Eq.~(\ref{1}) contains nothing explicitly about the surface properties signified by $\mathbf{F}_s$. This means that $g(\mathbf{x},\mathbf{v},t)$ at best gives the electronic distribution for the bulk of the metal. It does not apply to the surface layer. In particular, in contrast to what one would usually think~\cite{harris1972,garcia1977}, the distribution should not be used to compute the charge density in the surface layer according to the following formula,
$$\tilde{\rho}(\mathbf{x},t) = \left(\frac{m}{2\pi \hbar}\right)^2 \int d^3\mathbf{v} ~ eg(\mathbf{x},\mathbf{v},t),$$ 
where $\hbar$ is the reduced Planck constant. This point should also be clear in the absence of $V_s$ from $f_0\left(\varepsilon(\mathbf{v})\right)$, which implies again that Eq.~(\ref{1}) holds only in the bulk. We shall discuss this point further in the next section, where a more complete semi-classical method is presented to explicitly take care of the surface effects. 

As $\mathbf{F}_s$ is not present, Eq.~(\ref{1}) is translationally invariant along the surface and one can do a Fourier transform with respect to $\mathbf{r}$ and $t$ to find
\begin{equation}
\left(\partial_z - i\frac{\tilde{\omega}}{v_z}\right)g_{\mathbf{k}\omega}(z,\mathbf{v}) = -\frac{e \mathbf{v}\cdot\mathbf{E}_{\mathbf{k}\omega}(z)}{v_z} f'_0\left(\varepsilon \right), \label{2}
\end{equation}
where $f'_0 = \partial_\varepsilon f_0$, $\tilde{\omega} = \bar{\omega} - \mathbf{k}\cdot\mathbf{v}_\parallel$ with $\bar{\omega} = \omega + i/\tau$, $\mathbf{k}$ and $\omega$ are the planar wave vector and the frequency respectively, $g_{\mathbf{k}\omega}$ is the corresponding Fourier component of $g$ and so is $\mathbf{E}_{\mathbf{k}\omega}$ for $\mathbf{E}$. As a first-order linear differential equation, Eq.~(\ref{2}) admits of the following general solution
\begin{equation}
g_{\mathbf{k}\omega}(z,\mathbf{v}) = e^{i\frac{\tilde{\omega}z}{v_z}}\left(C(\mathbf{v}) - \frac{e f'_0 \mathbf{v}}{v_z} \cdot \int^z_0 dz' ~ e^{-i\frac{\tilde{\omega}z'}{v_z}} \mathbf{E}_{\mathbf{k}\omega}(z') \right)
\end{equation}
where $C(\mathbf{v}) = g_{\mathbf{k}\omega}(0,\mathbf{v})$ comes as a parameter independent of $z$ to be determined by boundary conditions. Now requiring that $g_{\mathbf{k}\omega}$ vanish far away from the surface, $z\rightarrow\infty$, one obtains
\begin{equation}
g_{\mathbf{k}\omega}(z,\mathbf{v}) =  \frac{e f'_0 \mathbf{v}}{v_z}  \cdot \int^\infty_z dz' ~ e^{i\frac{\tilde{\omega}(z-z')}{v_z}} \mathbf{E}_{\mathbf{k}\omega}(z'), \quad \text{for} ~ v_z<0. \label{4}
\end{equation}
Following Fuchs~\cite{abrikosov,ziman,fuchs1938,sondheimer1948,sondheimer1952,kaganov1997}, one assumes that at $z=0$ a fraction $p$ of incident electrons are specularly reflected back, namely
$$g_{\mathbf{k}\omega}(0,\mathbf{v}) = p g_{\mathbf{k}\omega}(0,\mathbf{v}_-), \quad \text{for} ~ v_z\geq0,$$ 
where $\mathbf{v}_- = (\mathbf{v}_\parallel, - v_z)$. This leads to
\begin{eqnarray}
&~& g_{\mathbf{k}\omega}(z,\mathbf{v}) = - p \frac{e f'_0 \mathbf{v}_-}{v_z}  \cdot \int^\infty_0 dz' ~ e^{i\frac{\tilde{\omega}(z+z')}{v_z}} \mathbf{E}_{\mathbf{k}\omega}(z')  \label{5} \\
 &~& \quad \quad  \quad \quad - \frac{e f'_0 \mathbf{v}}{v_z}  \cdot \int^z_0 dz' ~ e^{i\frac{\tilde{\omega}(z-z')}{v_z}} \mathbf{E}_{\mathbf{k}\omega}(z'), \quad \text{for} ~ v_z\geq0. \nonumber
\end{eqnarray}
Equations (\ref{4}) and (\ref{5}) constitute the basis of the Fuchs-Sondheimer method. One should see that, although it is usually taken as a simple constant, $p$ is allowed to vary with $\mathbf{k}$ and $\omega$ as well as $\mathbf{v}$. While people usually interpret $p$ after Fuchs, its exact physical meaning is not clear in the method. We shall show in Sec.~\ref{sec:3.2} that the Fuchs interpretation is to a large extent consistent with the Kubo-Greenwood formula but with some interesting discrepancies. 

The Fourier components of the current density are given by
$$\mathbf{J}_{\mathbf{k}\omega}(z) = \left(\frac{m}{2\pi \hbar}\right)^2 \int d^3\mathbf{v} ~ e \mathbf{v} g_{\mathbf{k}\omega}(z,\mathbf{v}).$$ We emphasize again that the as-obtained current density $\mathbf{J}$ applies only to the bulk of the metal. It does not apply to the surface region. As a notice of notation, we hereafter always use $\mathbf{J}$ for the bulk while reserving $\mathbf{j}$ for the entire system.
Now the conductivity tensor is obtained as
\begin{equation}
\sigma^{\mu\nu}_{\mathbf{k}\omega}(z,z') = e^2\left(\frac{m}{2\pi \hbar}\right)^2 \int_> d^3\mathbf{v} \left(-f'_0(\varepsilon)\right) \frac{v_\mu \Gamma^{\mu\nu}_{\mathbf{k}\omega}(z,z')v_\nu}{v_z}, \label{6}
\end{equation}
where the integral is restricted to $v_z\geq0$ as indicated by the symbol ``$>$", and the matrix $\Gamma$ can be obtained by direct substitution, given by
\begin{widetext}
\begin{equation}
\Gamma^{\mu\nu}_{\mathbf{k}\omega}(z,z') = 
\begin{cases}
\alpha\Theta(z-z') e^{-i\frac{\tilde{\omega}}{v_z}(z'-z)} + \Theta(z'-z) e^{i\frac{\tilde{\omega}}{v_z}(z'-z)} + p e^{i\frac{\tilde{\omega}}{v_z}(z'+z)}, & \text{if} ~~ \mu=x,y; \nu = x,y \\
\alpha\Theta(z-z') e^{-i\frac{\tilde{\omega}}{v_z}(z'-z)} + \Theta(z'-z) e^{i\frac{\tilde{\omega}}{v_z}(z'-z)} - p e^{i\frac{\tilde{\omega}}{v_z}(z'+z)}, & \text{if} ~~ \mu = z; \nu=z \\
\alpha\Theta(z-z') e^{-i\frac{\tilde{\omega}}{v_z}(z'-z)} - \Theta(z'-z) e^{i\frac{\tilde{\omega}}{v_z}(z'-z)} - p e^{i\frac{\tilde{\omega}}{v_z}(z'+z)}, & \text{if} ~~ \mu = x,y; \nu=z \\
\alpha\Theta(z-z') e^{-i\frac{\tilde{\omega}}{v_z}(z'-z)} - \Theta(z'-z) e^{i\frac{\tilde{\omega}}{v_z}(z'-z)} + p e^{i\frac{\tilde{\omega}}{v_z}(z'+z)}, & \text{if} ~~ \mu = z; \nu=x,y.
\end{cases}
\label{7}
\end{equation} 
\end{widetext}
Here $\alpha = 1$. As to be shown in Sec.~\ref{sec:3.2}, the same conductivity tensor as Eq.~(\ref{6}) obtains in the semi-classical limit of the Kubo-Greenwood formula, together with the matrix $\Gamma$, the main differences being that, by Kubo-Greenwood method $\alpha$ should be smaller than unity except for perfectly smooth surfaces, and in general one needs two specularity parameters to fully specify the surface effects. 

For the sake of completeness, we note down the following obvious relation
$$J^\mu_{\mathbf{k}\omega}(z) = \sum_{\nu=x,y,z}\int^\infty_0 dz' ~ \sigma^{\mu\nu}_{\mathbf{k}\omega}(z,z') E^\nu_{\mathbf{k}\omega}(z').$$
Conforming to the translational symmetry inherent in the method, this relation shows that a component $\mathbf{E}_\mathbf{k}$ can only induce a current with the same wave vector $\mathbf{k}$.

\section{Conductivity of metals bounded by a rough surface}
\label{sec:3}
As explained in the preceding section, the translational symmetry along an arbitrary surface is inherently built in the Fuchs-Sondheimer method. This is because the method does not explicitly include the confining surface potential $V_s$. In this section, we present calculations to fill this gap. We do this in two complementary approaches, the first (Sec.~\ref{sec:3.1}) based on a sort of generalized Fuchs-Sondheimer method with $V_s$ explicitly included in Boltzmann's equation, and the second (Sec.~(\ref{sec:3.2})) based on the quantal Kubo-Greenwood formula. We show that for a macroscopically flat surface, translational symmetry is indeed respected. 

As $V_s$ is fully considered in both approaches, we naturally arrive at two microscopic definitions, a semi-classical one by Eq.~(\ref{15}) and a quantum mechanical one by Eq.~(\ref{43}) for the specularity parameter. So far as we can see, the semi-classical definition (\ref{15}) does not lend itself to support the Fuchs interpretation. Indeed, there does not seem to exist any straightforward way to interpret (\ref{15}) as a fraction of specularly reflected electrons. On the other hand, the quantum mechanical definition (\ref{43}) does support Fuchs interpretation to a certain extent but with important differences. Firstly, according to Eq.~(\ref{43}), the specularity parameter counts more than specularly electrons: diffusively scattered electrons can also contribute. Secondly, instead of one parameter, two parameters are needed to fully specify the surface effects, as a result of the anisotropy in the system: the presence of a surface makes the planar direction distinct from the normal one. In spite of this, we find that in response to a uniform electric field the Fuchs interpretation is fully vindicated. In such cases, the two parameters become equal and do give the fraction of specularly reflected electrons, as Fuchs originally proposed. 

We shall also propose an extrapolation for obtaining the current density $\mathbf{j}$ valid for the entire system from that for the bulk region, the latter denoted by $\mathbf{J}$. While the extrapolation itself is simple, its generality has not been always recognized in the literature. 

\subsection{Calculations by generalized Fuchs-Sondheimer method}
\label{sec:3.1}
The basic equation used in the Fuchs-Sondheimer method, i.e. Eq.~(\ref{1}) does not include the surface confining potential $V_s(\mathbf{x})$ as the origin of surface roughness. Here we generalize it by including the latter explicitly. Again the electronic distribution splits into an equilibrium part $f_0\left(\varepsilon(\mathbf{x},\mathbf{v})\right)$, which is the Dirac-Fermi function as before, and a non-equilibrium part $g(\mathbf{x},\mathbf{v},t)$ that is induced by an electric field $\mathbf{E}(\mathbf{x},t)$. Note that $f_0$ depends on both $\mathbf{v}$ and $\mathbf{x}$ via $\varepsilon(\mathbf{x},\mathbf{v}) = V_s(\mathbf{x}) + m\mathbf{v}^2/2$. As a consequence of the profile of $V_s$, $f_0$ is the same as that in the Fuchs-Sondheimer method in the bulk region but diminishes in the surface region and eventually vanishes in the vacuum. The generalized Boltzmann equation reads~\cite{note}
\begin{equation}
\left(\mathcal{D}_t + \mathbf{v}\cdot \partial_\mathbf{x} + \frac{\mathbf{F}_s(\mathbf{x})}{m}\cdot\partial_\mathbf{v}\right)g(\mathbf{x},\mathbf{v},t) = - \frac{e\mathbf{E}(\mathbf{x},t)}{m}\cdot \partial_{\mathbf{v}} f_0\left(\varepsilon(\mathbf{x},\mathbf{v})\right). \nonumber
\end{equation}
As this equation includes surface effects in totality, the resulting distribution $g(\mathbf{x},\mathbf{v},t)$ applies to the entire system and can be used to calculate both the charge density $\rho(\mathbf{x},t)$ and the current density $\mathbf{j}(\mathbf{x},t)$ everywhere including the surface region. 

\begin{widetext}
After a Fourier transform to the above equation, we find
\begin{equation}
\left(\partial_z - i\frac{\tilde{\omega}}{v_z}\right)g_{\mathbf{k}\omega}(z,\mathbf{v}) = -\frac{1}{m v_z} \sum_{\mathbf{k}'} \mathbf{F}^{\mathbf{k}-\mathbf{k}'}_s(z) \cdot\partial_\mathbf{v} g_{\mathbf{k}'\omega}(z,\mathbf{v}) - \frac{e \mathbf{v}}{v_z} \cdot \sum_{\mathbf{k}'} \mathbf{E}_{\mathbf{k}-\mathbf{k}'\omega}(z)f'_{0,\mathbf{k}'}\left(z,\mathbf{v}\right), \label{8}
\end{equation}
where $f'_{0,\mathbf{k}}(z,\mathbf{v})$ is the Fourier transform of $f'_0\left(\varepsilon(\mathbf{x},\mathbf{v})\right)$ with respect to $\mathbf{r}$. Equation (\ref{8}) reduces to Eq.~(\ref{1}) in the bulk region. It can be rewritten in an integral form,
\begin{equation}
g_{\mathbf{k}}(z,\mathbf{v}) = g^{(0)}_\mathbf{k}(z,\mathbf{v}) - \frac{1}{mv_z} \sum_{\mathbf{k}'} \int^\infty_{-\infty} dz' G_{\mathbf{k}\mathbf{v}}(z-z') \mathbf{F}^{\mathbf{k}-\mathbf{k}'}_s(z')\cdot\partial_\mathbf{v}g_{\mathbf{k}'}(z',\mathbf{v}) - \frac{e \mathbf{v}}{v_z} \cdot \int^\infty_{-\infty} dz' G_{\mathbf{k}\mathbf{v}}(z-z') \mathbf{E}_{\mathbf{k}-\mathbf{k}'}(z') f'_{0,\mathbf{k}'}(z',\mathbf{v}).  \label{9}
\end{equation}
Here we have omitted the subscript $\omega$ to simplify the notation, and $g^{(0)}$ and $G$ are defined by
\begin{equation}
\left(\partial_z - i\frac{\tilde{\omega}}{v_z}\right)g^{(0)}_{\mathbf{k}}(z,\mathbf{v}) = 0, \quad \left(\partial_z - i\frac{\tilde{\omega}}{v_z}\right)G_{\mathbf{k}\mathbf{v}}(z-z') = \delta(z-z'),
\end{equation}
where $\delta(z)$ is the Dirac delta function. These equations are easily solved to yield 
$$g^{(0)}_{\mathbf{k}} = A_\mathbf{k}(\mathbf{v}) e^{i\frac{\tilde{\omega} z}{v_z}}, \quad G_{\mathbf{k}\mathbf{v}}(z-z') = \left(\Theta(z-z')a_\mathbf{k}(\mathbf{v}) + \Theta(z'-z)b_\mathbf{k}(\mathbf{v})\right)e^{i\frac{\tilde{\omega} (z-z')}{v_z}},$$ 
where $A_\mathbf{k}(\mathbf{v})$, $a_\mathbf{k}(\mathbf{v})$ and $b_\mathbf{k}(\mathbf{v}) = a_\mathbf{k}(\mathbf{v}) - 1$ are parameters. By requiring that $g_\mathbf{k}(z,\mathbf{v})$ vanish for $z\rightarrow\pm\infty$, these parameters can be uniquely determined and we arrive at
\begin{equation}
g_{\mathbf{k}}(z,\mathbf{v}) = \frac{1}{v_z} \int^\infty_{-\infty} dz' ~ \left(\Theta(-v_z)\Theta(z'-z) - \Theta(v_z)\Theta(z-z') \right) e^{i\frac{\tilde{\omega}(z-z')}{v_z}}  \sum_{\mathbf{k}'} \left(e\mathbf{v}\cdot \mathbf{E}_{\mathbf{k}-\mathbf{k}'}(z')f'_{0,\mathbf{k}'}(z',\mathbf{v}) + \frac{\mathbf{F}^{\mathbf{k}-\mathbf{k}'}_s(z')}{m}\cdot\partial_\mathbf{v}g_{\mathbf{k}'}(z',\mathbf{v})\right). \label{11}
\end{equation}
\end{widetext}
The full solution to this equation can in principle be obtained by means of iteration, though we do not really need it for the current purpose. The planar translational symmetry is broken due to the terms in the sum, which allows a given Fourier component $\mathbf{E}_{\mathbf{q}}$ to generate a spectrum of $g_{\mathbf{k}}$ with $\mathbf{k}$ differing from $\mathbf{q}$, that is, the current density has non-vanishing components with wave vectors other than those of the electric field. Note that Eq.~(\ref{11}) puts the distribution for electrons with $v_z\geq0$ and those with $v_z<0$ on formally equal footing, unlike in the Fuchs-Sondheimer method. 

To gain some insight into $g_{\mathbf{k}}(z,\mathbf{v})$, let us have a look at its behaviors in the vacuum side, where $z<-z_s/2$. Note that for $v_z\geq0$ the integral runs over $z'\in (-\infty,z]$. Physically, this is because electrons with $v_z\geq0$ travel toward $z$ from $z'<z$ and thus only feel the impact of the field in that interval. In this interval, however, $f_0$ vanishes and so does $\mathbf{F}_s$. As a result, the integral in Eq.~(\ref{11}) also vanishes, returning $g_{\mathbf{k}}(z,\mathbf{v}) = 0$ for $v_z\geq0$ in the vacuum. This is, of course, consistent with the fact that electrons are confined to the metal. For $v_z<0$, the integral runs over $z'\in [z,\infty)$ as this involves electrons traveling from $z'>z$. Considering that $\mathbf{F}_s$ is discernible only in the surface region, we may take
\begin{equation}
\int^\infty_{z<-\frac{z_s}{2}} dz' ~\mathbf{F}^{\mathbf{k}-\mathbf{k}'}_s(z')\cdot\partial_\mathbf{v} g_{\mathbf{k}'}(z',\mathbf{v}) \approx z_s \mathbf{F}^{\mathbf{k}-\mathbf{k}'}_s(0)\cdot\partial_\mathbf{v} g_{\mathbf{k}'}(0,\mathbf{v}). \nonumber
\end{equation}
On the other hand, as electrons are confined to the metal, $g_{\mathbf{k}}(z,\mathbf{v})$ has to vanish in the vacuum also for $v_z<0$. To conform to this fact, we are left to conclude that
\begin{widetext}
\begin{equation}
\frac{z_s}{m} \sum_{\mathbf{k}'}\mathbf{F}^{\mathbf{k}-\mathbf{k}'}_s(0)\cdot\partial_\mathbf{v} g_{\mathbf{k}'}(0,\mathbf{v}) \approx - \sum_{\mathbf{k}'} \int^\infty_{-z_s/2} dz' ~ e^{-i\frac{\tilde{\omega}z'}{v_z}} e\mathbf{v}\cdot \mathbf{E}_{\mathbf{k}-\mathbf{k}'}(z') f'_{0,\mathbf{k}'}(z',\mathbf{v}), \quad v_z<0. \label{12}
\end{equation}
Here the integral over $[z<-z_s/2,\infty)$ has been replaced with one over $[-z_s/2,\infty)$, because $f_0$ is negligible in the vacuum. 

We proceed to look at $g_{\mathbf{k}}(z,\mathbf{v})$ in the bulk of the metal, where $z>z_s/2$. By a similar line of reasoning, one obtains
\begin{equation}
g_{\mathbf{k}}\left(z>\frac{z_s}{2},\mathbf{v}\right) = g^{(b)}_{\mathbf{k}}(z,\mathbf{v}) = 
\begin{cases}
\frac{ef'_0\left(\varepsilon(\mathbf{v})\right) \mathbf{v}}{v_z} \cdot \int^\infty_z dz'~ e^{i\frac{\tilde{\omega}(z-z')}{v_z}} \mathbf{E}_{\mathbf{k}}(z'), & \text{for}~v_z<0 \\
- \frac{e \mathbf{v}}{v_z} \cdot \sum_{\mathbf{k}'} \int^z_{-\frac{z_s}{2}} dz' ~ e^{i\frac{\tilde{\omega}(z-z')}{v_z}} \mathbf{E}_{\mathbf{k}-\mathbf{k}'}(z') f'_{0,\mathbf{k}'}(z',\mathbf{v}) - e^{i\frac{\tilde{\omega}z}{v_z}}\left[\frac{z_s}{mv_z} \sum_{\mathbf{k}'} \mathbf{F}^{\mathbf{k}-\mathbf{k}'}_s(0)\cdot\partial_\mathbf{v} g_{\mathbf{k}'}(0,\mathbf{v})\right], & \text{for}~v_z\geq0 
\end{cases}
\label{13}
\end{equation} 
\end{widetext}
Here in the upper line for $v_z<0$, we have considered that $f_0\left(\varepsilon(\mathbf{x},\mathbf{v})\right)$ reduces to $f_0\left(\varepsilon(\mathbf{v})\right)$ in the bulk, and thus $f'_{0,\mathbf{k}} = \delta_{\mathbf{k},0} f'_0\left(\varepsilon(\mathbf{v})\right)$. It is clear that the surface does not affect the distribution of electrons with $v_z<0$ in the bulk, as these electrons have not reached the surface yet in the regime of linear response, and thus $g$ retains the planar translational symmetry for them regardless of surface roughness. For electrons with $v_z\geq0$, the distribution is inevitably affected by the surface and the planar translational symmetry is lost. Both terms in the lower line of Eq.~(\ref{13}) break the symmetry. In the first term, the integral can be split as $\int^z_{-z_s/2} = \int^z_{z_s/2} + \int^{z_s/2}_{-z_s/2}$. Only the surface contribution from $\int^{z_s/2}_{-z_s/2}$ breaks the symmetry. 

However, if the surface is macroscopically flat, the symmetry can be restored. By ``macroscopic flatness", we mean that $V^{\mathbf{k}}_s(z)$ (the Fourier components of $V_s(\mathbf{x})$) is negligible unless $k\approx0$ or $k$ is comparable to the inverse of a lattice constant $a^{-1} \sim  k_F$, where $k = \abs{\mathbf{k}}$ and $k_F$ is the electronic Fermi wavelength. We further restrict ourselves to considering electric field that does not vary significantly over the scale $1/k_F$. Under such circumstances, we have $\sum_{\mathbf{k}'}\mathbf{E}_{\mathbf{k}-\mathbf{k}'}(z')f'_{0,\mathbf{k}'}(z') \approx \mathbf{E}_\mathbf{k}(z') f'_{0,0}(z',\mathbf{v})$ for $\mathbf{k}$ with $k$ much smaller than $k_F$, as is indeed the case for coarse-grained quantities~\cite{note2}. Substituting this into the lower line of Eq.~(\ref{13}), the distribution for $v_z\geq0$ \textit{in the bulk} becomes~\cite{note3}
\begin{eqnarray}
g^{(b)}_{\mathbf{k}}(z,\mathbf{v}) &=& - \frac{e  f'_{0}\left(\varepsilon(\mathbf{v})\right)\mathbf{v}}{v_z} \cdot \int^z_{-z_s/2} dz' ~ e^{i\frac{\tilde{\omega}(z-z')}{v_z}} \mathbf{E}_{\mathbf{k}}(z')  \nonumber \\ &~& ~ ~ ~ ~ - e^{i\frac{\tilde{\omega}z}{v_z}}\left[\frac{z_s}{mv_z} \sum_{\mathbf{k}'} \mathbf{F}^{\mathbf{k}-\mathbf{k}'}_s(0)\cdot\partial_\mathbf{v} g_{\mathbf{k}'}(0,\mathbf{v})\right]. 
\label{14}
\end{eqnarray}
Here the first term obviously respects the symmetry, as expected. Now the second term also respects the symmetry. To show this, let us observe that, under ``macroscopic flatness", the components $g_{\mathbf{k}'}$ in the sum over $\mathbf{k}'$ in this term have either $k'\sim k$ (specular reflection) or $k' \sim k_F$ (diffusive reflection), provided $k$ is small as is so for coarse-grained quantities. The specular reflection part preserves the symmerty, while the diffusive part involves only short-wavelength components and does not mix with long-wavelength components of $g$. As such, Eq.~(\ref{14}) can be solved for $g_\mathbf{k}$ with small $k$ separately from those with large $k$. The components $g_\mathbf{k}$ with large $k$ are irrelevant after coarse-graining. We can thus define the following parameter, 
\begin{equation}
p_{\mathbf{k}} = \frac{\frac{1}{v_z} \sum_{\mathbf{k}'} \mathbf{F}^{\mathbf{k}-\mathbf{k}'}_s(0)\cdot\partial_\mathbf{v} g_{\mathbf{k}'}(0,\mathbf{v})}{\frac{1}{-v_z} \sum_{\mathbf{k}'} \mathbf{F}^{\mathbf{k}-\mathbf{k}'}_s(0)\cdot\partial_{\mathbf{v}_-} g_{\mathbf{k}'}(0,\mathbf{v}_-)}, \quad v_z\geq0. \label{15}
\end{equation}
Now, combined with Eq.~(\ref{12}), this relation ensures that the second term in Eq.~(\ref{14}) also respects the symmetry. In this way, the translational symmetry along a rough surface is restored for coarse-grained quantities. Indeed, if $p_\mathbf{k}$ is identified with the specularity parameter introduced in the Fuchs-Sondheimer method, Eq.~(\ref{14}) becomes the same as the distribution Eq.~(\ref{5}) in the macroscopic limit $z_s/\Lambda\ll1$. 

Equation (\ref{15}) may then serve as a microscopic definition for the specularity parameter. Obviously, the expression reflects the way the surface impacts discriminately the distribution of incoming and departing electrons. If the surface is perfectly smooth, i.e. in the absence of diffusive scattering so that $\mathbf{F}^{\mathbf{k}-\mathbf{k}'}_s(z) = \delta_{\mathbf{k},\mathbf{k}'} \left(0,0,F_s(z)\right)$, then $p_\mathbf{k} = \frac{\partial_{v_z}g_{\mathbf{k}}(0,\mathbf{v})}{\partial_{v_z}g_{\mathbf{k}}(0,\mathbf{v}_-)}$ may well be unity provided that $g_\mathbf{k}$ is symmetric about incoming and reflected back electrons. However, it is unclear to us how $p_\mathbf{k}$ can be simply identified with the fraction of electrons that are specularly reflected at the surface. Additionally, its dependence on $\mathbf{v}$ disadvantages it as a proper intrinsic characterization of surface roughness, as remarked before.

While it acquires a simple form in terms of the specularity parameter without explicitly referring to the particulars of the surface confining potential $V_s$, $g^{(b)}_\mathbf{k}$ is valid only in the bulk region. On the other hand, the global distribution $g_\mathbf{k}$, as given by Eq.~(\ref{11}), is valid in the entire system, but it explicitly involves $V_s$. Because of this, the global current density $\mathbf{j}$, which is calculated with $g$ in the usual way and applies to the entire system, may prove inconvenient for use since $V_s$ is almost impossible to be experimentally studied. It is therefore desirable to relate $\mathbf{j}$ to its bulk form, $\mathbf{J}$ based on $g^{(b)}$ and only valid in the bulk. To this end, one may introduce a phenomenological function $w_\mu(z)$ to relate $j_\mu$ and $J_\mu$. Namely, we write $j_{\mu,\mathbf{k}}(z) = w_\mu(z) J_{\mu,\mathbf{k}}(z)$. By definition, $w_\mu(z)$ must approach unity for $z>z_s/2$ and zero for $z<-z_s/2$. In addition, it should be a smooth function across the surface region. Its microscopic profile certainly depends on $V_s$ and can in principle be determined from Eq.~(\ref{11}) but is not interesting to us. Instead, we observe that, on the macroscopic scale of $\Lambda\gg z_s$, the details of $w_\mu(z)$ are irrelevant and this function degenerates with the step function $\Theta(z)$ and therefore becomes independent of $\mathbf{k}$, $\mathbf{v}$ and $V_s$. This allows one to obtain $\mathbf{j}_\mathbf{k}(z)$ from $\mathbf{J}_\mathbf{k}(z)$ as follows,
\begin{equation}
\mathbf{j}_\mathbf{k}(z) = \Theta(z) \mathbf{J}_\mathbf{k}(z). \label{16}
\end{equation}
Here it is implicitly understood that both $\mathbf{j}$ and $\mathbf{J}$ are coarse-grained on the scale of $\Lambda$. Further, by means of the equation of continuity, the global charge density $\rho$, which is valid for the entire system and can certainly be calculated with $g$ [c.f. Eq.~(\ref{11})]~\cite{note4}, can be calculated from Eq.~(\ref{16}), as recently demonstrated in Refs.~\cite{deng2019,deng2018}. 

The Fuchs-Sondheimer method allows one to calculate $\mathbf{J}$ but not $\mathbf{j}$. Though this is a simple observation, some existing work have failed to distinguish them and drawn incorrect conclusions, see Sec.~\ref{sec:4}. 

\subsection{Calculations by Kubo-Greenwood method}
\label{sec:3.2}
We proceed to examine the conductivity on the basis of the Kubo-Greenwood formula and further clarify the physical meaning of the parameter $p$. It will be shown that, in this formalism the conductivity can also be obtained in closed form even if the microscopic details of the surface profile are unknown, as much as  with the Fuchs-Sondheimer method. The role of $p$ now has to be played by a surface scattering matrix, which in the semi-classical limit reduces to a pair of parameters denoted by $p_\parallel$ and $p_z$, respectively. 

The unperturbed single-particle Hamiltonian of the electron gas is given by
$$H = -\left(\hbar^2/2m\right)\partial^2_\mathbf{x} + V_s(\mathbf{x}).$$ We shall assume that $V_s$ is purely repulsive and no bound states can form of $H$. The entire system is put in an imaginary huge cubic box with linear dimension $L$, face area $A = L^2$ and volume $\mathcal{V} = L^3$, for the convenience of normalization. Thus, any stationary state of $H$ can be written as a scattering state $|\psi_{\mathbf{k}q}\rangle$ with the following wave function,
\begin{equation}
\psi_{\mathbf{k}q}(\mathbf{x}) = \frac{e^{i(\mathbf{k}\cdot\mathbf{r}-qz)}}{\sqrt{\mathcal{V}}} + \psi^s_{\mathbf{k}q}(\mathbf{x}), \quad q\geq0,
\end{equation}
where the first term represents a wave incident on the surface with wave vector $\mathbf{K} = (\mathbf{k},-q)$ of magnitude $K = \sqrt{k^2+q^2}$, and $\psi^s_{\mathbf{k}q}$ represents the scattered waves, which can be obtained using Lippman-Schwinger formula as
\begin{equation}
\psi^s_{\mathbf{k}q}(\mathbf{x}) = \sum_{\mathbf{k}_s} \frac{e^{i\mathbf{k}_s\cdot\mathbf{r}}}{\sqrt{\mathcal{V}}} \int^\infty_{-\infty} dz' ~ G_{K}(z-z';\mathbf{k}_s) \langle \mathbf{k}_s,z'|\hat{T}|\mathbf{k},q\rangle. \label{18}
\end{equation}
Here $G_{K}(z-z';\mathbf{k})$ is the Green's function solving 
$$\left(\partial^2_z + K^2-k^{2}\right)G_{K}(z-z';\mathbf{k}) = \delta(z-z'),$$ 
$\hat{T} = \tilde{V}_s\left(1-G_K\tilde{V}_s\right)^{-1} = \tilde{V}_s + \tilde{V}_sG_K\tilde{V}_s + ...$ denotes the transition matrix with $\tilde{V}_s = 2mV_s/\hbar^2$, and $|\mathbf{k},z\rangle = |\mathbf{k}\rangle \otimes |z\rangle$. The energy of $|\psi_{\mathbf{kq}}\rangle$ is written as $\varepsilon_{\mathbf{k}q} = \hbar^2K^2/2m$. The domain of integral in Eq.~(\ref{18}) can practically be restricted to the surface region, as $\hat{T}$ acts in this region only.

Let us reorganize the waves in $\psi_{\mathbf{k}q}$ according to 
\begin{equation}
\psi_{\mathbf{k}q} = \Psi_{\mathbf{k}q} + \psi^d_{\mathbf{k}q}, \label{psi}
\end{equation}
where $\Psi$ includes the incident wave and the specularly reflected wave with $\mathbf{k}_s = \mathbf{k}$, i.e.
\begin{equation}
\Psi_{\mathbf{k}q}(\mathbf{x}) = \frac{e^{i\mathbf{k}\cdot\mathbf{r}}}{\sqrt{\mathcal{V}}} \Phi_{\mathbf{k}q}(z), \quad \Phi_{\mathbf{k}q}(z) = e^{-iqz} + R_{\mathbf{k}q}(z),
\end{equation}
and $\psi^d_{\mathbf{k}q}$ collects all the diffusively scattered waves with $\mathbf{k}_s \neq \mathbf{k}$, which may be written as
\begin{equation}
\psi^d_{\mathbf{k}q}(\mathbf{x}) = \frac{e^{i\mathbf{k}\cdot\mathbf{r}}}{\sqrt{\mathcal{V}}} \phi_{\mathbf{k}q}(\mathbf{r},z).
\end{equation}
In the above, $R_{\mathbf{k}q}(z)$ and $\phi_{\mathbf{k}q}(\mathbf{r},z)$ are defined via Eq.~(\ref{18}) in a straightforward manner, the expressions of which we do not bother to write down. Nevertheless, we wish to point out that 
\begin{equation}
\phi_{\mathbf{k}q}(\mathbf{r},z) = \sum_{\mathbf{k}_d} \phi^{\mathbf{k}_d}_{\mathbf{k}q}(z) e^{i\mathbf{k}_d\cdot\mathbf{r}}, \label{phi} 
\end{equation}
where $\mathbf{k}_d = \mathbf{k}_s - \mathbf{k}$. Under the condition of ``macroscopic flatness", $k_d = \abs{\mathbf{k}_d}$ is of the order of $1/a \sim k_F$, which means that $\phi_{\mathbf{k}q}(\mathbf{r},z)$ varies rapidly with $\mathbf{r}$. We shall show that, due to this rapid variation, the interference between $\psi^d$ and $\Psi$ does not contribute to the electrical responses to an electric field that varies insignificantly over a planar distance $\sim 1/k_d$. 

Physical causality requires that 
\begin{equation}
G_K(z-z';\mathbf{k}) = \frac{1}{2i\sqrt{K^2-k^2}} e^{i\sqrt{K^2-k^2}\abs{z-z'}},
\end{equation}
so that in the bulk of the metal only the incident and reflected waves exist while in the vacuum only transmitted waves exist, the latter assumed to be negligible in accord with the fact that the electrons are confined to the metal by $V_s$. Considering that the impact of $\hat{T}$ is limited to the surface region, we obtain in the bulk region that
\begin{equation}
\psi^s_{\mathbf{k}q}(\mathbf{x}) = \sum_{\mathbf{k}_s} S_K(\mathbf{k},\mathbf{k}_s) \frac{e^{i(\mathbf{k}_s\cdot\mathbf{r} + q_sz)}}{\sqrt{\mathcal{V}}}, \quad \text{for} ~ z>\frac{z_s}{2},
\end{equation}
where $q_s = \sqrt{K^2-k^2_s}$ and $S_K$ is the scattering matrix,
\begin{equation}
S_K(\mathbf{k},\mathbf{k}_s) = \int_\text{surface region} dz ~ \frac{e^{-iq_sz}}{2iq_s} \langle\mathbf{k}_s,z|\hat{T}|\mathbf{k},q\rangle. \label{scaM}
\end{equation}
It follows that, in the bulk region
\begin{equation}
R_{\mathbf{k}q}(z) = R_{\mathbf{k}q} e^{iqz}, \quad \phi^{\mathbf{k}_d}_{\mathbf{k}q}(z) = S_K(\mathbf{k},\mathbf{k}+\mathbf{k}_d) e^{iq_sz}. \label{23}
\end{equation}
Here $R_{\mathbf{k}q}$ is the specular reflection amplitude. We see that the surface effects in the bulk region can be completely specified by the matrix $S_K$, which is reminiscent of the specularity parameter $p$ that is introduced to specify surface effects in the Fuchs-Sondheimer method.  

We are now ready to calculate the electrical conductivity. According to the Kubo-Greenwood formula, it is given by~\cite{notekubo}
\begin{equation}
\sigma^{\mu\nu}_\omega(\mathbf{x},\mathbf{x}') = i\hbar \sum_{\mathbf{k}q,\mathbf{k}'q'} \frac{f_0(\varepsilon') - f_0(\varepsilon)}{\varepsilon - \varepsilon'} \frac{\hat{j}^\mu_{\mathbf{k}q,\mathbf{k}'q'}(\mathbf{x}) \hat{j}^\nu_{\mathbf{k}'q',\mathbf{k}q}(\mathbf{x}')}{\hbar\bar{\omega} + \varepsilon - \varepsilon'}, \label{24}
\end{equation}
where $\varepsilon = \varepsilon_{\mathbf{k}q}$ and $\varepsilon' = \varepsilon_{\mathbf{k}'q'}$ are shorthands, and $\hat{j}^{\mu}_{a,b}(\mathbf{x})$ are the matrix elements of the $\mu$-th component of the current density operator $\hat{j}_{\mu}(\mathbf{x})$, i.e.
\begin{equation}
\hat{j}^{\mu}_{a,b}(\mathbf{x}) = \frac{e\hbar}{2mi}\left[\psi^*_a(\mathbf{x})\partial_\mu \psi_b(\mathbf{x}) - \left(\partial_\mu\psi^*_a(\mathbf{x})\right) \psi_b(\mathbf{x})\right]. \label{25}
\end{equation} 
Physically, Eq.~(\ref{24}) describes the response at $\mathbf{x}$ to a field at $\mathbf{x}'$ via an electron-hole pair created at $\mathbf{x}'$ by the field and thence propagating to $\mathbf{x}$. The current density $\mathbf{j}_\omega(\mathbf{x})$ is related to the electric field $\mathbf{E}_\omega(\mathbf{x})$ by
\begin{equation}
j^\mu_\omega(\mathbf{x}) = \sum_\nu \int d^3\mathbf{x}' ~ \sigma^{\mu\nu}_\omega(\mathbf{x},\mathbf{x}') E^\nu_\omega(\mathbf{x}'). \label{26}
\end{equation}
It should be noted that $\sigma(\mathbf{x},\mathbf{x}')$ -- hereafter the subscript $\omega$ is dropped -- is negligible if either $\mathbf{x}$ or $\mathbf{x}'$ or both lie in the vacuum with $z<-z_s/2$, since the wave functions and hence the matrix elements (\ref{25}) are assumed negligible in the vacuum. As such, the domain of integral in (\ref{26}) can practically be restricted to the metal.  

\subsubsection{Restoration of planar translational symmetry}
The conductivity (\ref{24}) does not possess translational symmetry along the surface due to diffusively scattered waves. We now discuss how this symmetry might be restored. To this end, we observe that, at low temperatures, the dominant contributions in (\ref{24}) come from terms with $\varepsilon$ and $\varepsilon'$ close to each other. Let us write for these terms that $\mathbf{k}' = \mathbf{k} + \delta\mathbf{k}$ and $q' = q + \delta q$, where $\delta k = \abs{\delta \mathbf{k}}$ and $\delta q$ are small. Further, we notice that $\sigma$ in Eq.~(\ref{26}) can be replaced with its coarse-grained version, if the electric field varies appreciably only over a distance much larger than $\Lambda > 1/k_d$ and only coarse-grained current densities are concerned. We thus coarse-grain $\sigma(\mathbf{x},\mathbf{x}')$ over both planar coordinates $\mathbf{r}$ and $\mathbf{r}'$ on the scale of $\Lambda$. As is clear from Eq.~(\ref{24}), this coarse-graining boils down to a separate coarse-graining of the matrix elements $\hat{j}^\mu_{\mathbf{k}q,\mathbf{k}'q'}(\mathbf{x})$ over $\mathbf{r}$ and $\hat{j}^\nu_{\mathbf{k}'q',\mathbf{k}q}(\mathbf{x}')$ over $\mathbf{r}'$. Substituting $\psi$ [c.f. Eq.~(\ref{psi})] into Eq.~(\ref{25}), one sees that these matrix elements generally each consist of three types of terms: one involving only $\Psi$, another involving only $\psi^d$ and the interference term involving both $\Psi$ and $\psi^d$. The Fourier components of the interference term have wave vectors $\mathbf{k}_d + \delta\mathbf{k}$ and are all fast varying: they vanish upon coarse-graining. In this way, we arrive at
\begin{equation}
\overline{\hat{\mathbf{j}}_{\mathbf{k}q,\mathbf{k}'q'}(\mathbf{x})} = \frac{e^{i(\mathbf{k}'-\mathbf{k})\cdot\mathbf{r}}}{\mathcal{V}} \mathbf{j}_{\mathbf{k}q,\mathbf{k}'q'}(z), \label{30}
\end{equation}
where the over-line in $\overline{O}$ does the coarse-graining of $O$ and 
$$\mathbf{j}_{\mathbf{k}q,\mathbf{k}'q'}(z) = \mathbf{j}^{\Psi}_{\mathbf{k}q,\mathbf{k}'q'}(z) + \mathbf{j}^{d}_{\mathbf{k}q,\mathbf{k}'q'}(z),$$ with $\mathbf{j}^{\Psi}$ and $\mathbf{j}^{d}$ stemming from $\Psi$ and $\psi^d$, respectively. Direct substitution gives
\begin{equation}
\mathbf{j}^{\Psi}_{\mathbf{k}q,\mathbf{k}'q'}(z) = \frac{e\hbar}{2m} 
\begin{pmatrix}
(k_x+k'_x) \Phi^*_{\mathbf{k}q}(z)\Phi_{\mathbf{k}'q'}(z) \\
(k_y+k'_y) \Phi^*_{\mathbf{k}q}(z)\Phi_{\mathbf{k}'q'}(z) \\
\Phi^*_{\mathbf{k}q}(z)\frac{\partial_z}{i}\Phi_{\mathbf{k}'q'}(z) - \frac{\partial_z}{i}\Phi^*_{\mathbf{k}q}(z)\Phi_{\mathbf{k}'q'}(z)
\end{pmatrix}. \label{31}
\end{equation}
Before we give the expression of $\mathbf{j}^{d}_{\mathbf{k}q,\mathbf{k}'q'}(z)$, let us show that 
$$\overline{\phi^*_{\mathbf{k}q}(\mathbf{r},z)(-i\partial_\mathbf{r})\phi_{\mathbf{k}'q'}(\mathbf{r},z)} = 0.$$ Indeed, upon substituting Eq.~(\ref{phi}) for $\phi_{\mathbf{k}q}$ and using $$\overline{e^{i(\mathbf{k}_d-\mathbf{k}'_d)\cdot\mathbf{r}}} \approx \delta_{\mathbf{k}_d,\mathbf{k}'_d},$$ the expression in question becomes $\sum_{\mathbf{k}_d} \abs{\phi^{\mathbf{k}_d}_{\mathbf{k}q}(z)}^2 \mathbf{k}_d$, which vanishes if $\abs{\phi^{\mathbf{k}_d}_{\mathbf{k}q}(z)}^2$ depends only on the magnitude of $\mathbf{k}_d$ as is reasonable for a rough surface, thus completing the proof. Now we obtain
\begin{equation}
\mathbf{j}^{d}_{\mathbf{k}q,\mathbf{k}'q'}(z) = \frac{e\hbar}{2m}
\begin{pmatrix}
(k_x+k'_x) \mathcal{P}^\parallel_{\mathbf{k}q,\mathbf{k}'q'}(z) \\
(k_y+k'_y) \mathcal{P}^\parallel_{\mathbf{k}q,\mathbf{k}'q'}(z) \\
(q+q')\mathcal{P}^z_{\mathbf{k}q,\mathbf{k}'q'}(z)
\end{pmatrix}, \label{32}
\end{equation}
with $\mathcal{P}^\parallel_{\mathbf{k}q,\mathbf{k}'q'}(z) = \overline{\phi^*_{\mathbf{k}q}(\mathbf{r},z)\phi_{\mathbf{k}'q'}(\mathbf{r},z)} \approx \sum_{\mathbf{k}_d} \phi^{\mathbf{k}_d*}_{\mathbf{k}q}(z)\phi^{\mathbf{k}_d}_{\mathbf{k}'q'}(z)$ and 
\begin{widetext}
\begin{equation}
\mathcal{P}^z_{\mathbf{k}q,\mathbf{k}'q'}(z) = \frac{1}{q+q'}\left(\overline{\phi^*_{\mathbf{k}q}(\mathbf{r},z)\frac{\partial_z}{i}\phi_{\mathbf{k}'q'}(\mathbf{r},z)} - \overline{ \frac{\partial_z}{i}\phi^*_{\mathbf{k}q}(\mathbf{r},z)\phi_{\mathbf{k}'q'}(\mathbf{r},z)}\right) \approx \frac{1}{q+q'}\sum_{\mathbf{k}_d} \left(\phi^{\mathbf{k}_d*}_{\mathbf{k}q}(z)\frac{\partial_z}{i}\phi^{\mathbf{k}_d}_{\mathbf{k}'q'}(z) - \frac{\partial_z}{i} \phi^{\mathbf{k}_d*}_{\mathbf{k}q}(z)\phi^{\mathbf{k}_d}_{\mathbf{k}'q'}(z)\right). \label{33}
\end{equation}
The coarse-grained conductivity, which is still to be called $\sigma^{\mu\nu}(\mathbf{x},\mathbf{x}')$, can be obtained from Eq.~(\ref{24}) by replacing $\hat{j}^\mu_{\mathbf{k}q,\mathbf{k}'q'}(\mathbf{x})$ and $\hat{j}^\nu_{\mathbf{k}'q',\mathbf{k}q}(\mathbf{x}')$ with their respective coarse-grained counterparts [c.f. Eq.~(\ref{30})]. It now only depends on $\mathbf{r}-\mathbf{r}'$ not on $\mathbf{r}$ and $\mathbf{r}'$ separately any more, i.e. $\sigma^{\mu\nu}(\mathbf{x},\mathbf{x}') = \sigma^{\mu\nu}(\mathbf{r}-\mathbf{r}',z,z')$. The translational symmetry along a rough surface has thus been restored. Writing 
$ \sigma^{\mu\nu}(\mathbf{r}-\mathbf{r}',z,z') = \sum_{\delta\mathbf{k}} \sigma^{\mu\nu}_{\delta\mathbf{k}}(z,z') \frac{e^{i \delta\mathbf{k}\cdot(\mathbf{r}-\mathbf{r}')}}{A},$ we find the Fourier components as
\begin{equation}
\sigma^{\mu\nu}_{\delta\mathbf{k}}(z,z') = \frac{i\hbar}{\mathcal{V}}\sum_{\mathbf{k}q}\frac{1}{L}\sum_{\delta q}  \frac{f_0(\varepsilon_{\mathbf{k}+\delta\mathbf{k} q+\delta q}) - f_0(\varepsilon_{\mathbf{k}q})}{\varepsilon_{\mathbf{k}q} - \varepsilon_{\mathbf{k}+\delta\mathbf{k} q+\delta q}} \frac{j^\mu_{\mathbf{k}q,\mathbf{k}+\delta\mathbf{k} q+\delta q}(z) j^\nu_{\mathbf{k}+\delta\mathbf{k} q+\delta q,\mathbf{k}q}(z')}{\hbar\bar{\omega} + \varepsilon_{\mathbf{k}q} - \varepsilon_{\mathbf{k}+\delta\mathbf{k} q+\delta q}}, \quad j^\mu_{\delta \mathbf{k}}(z) = \sum_\nu \int dz'~\sigma^{\mu\nu}_{\delta \mathbf{k}}(z,z') E^\nu_{\delta \mathbf{k}}(z'). \label{34}
\end{equation}
This expression applies to the entire system, including both the bulk and the surface regions of the metal. As in the semi-classical method, for $z$ lying in the vacuum it vanishes. 

\subsubsection{Semi-classical limit and definition of \textit{Fuchs} parameter}
We proceed to examine $\sigma^{\mu\nu}_{\delta\mathbf{k}}(z,z')$ in the semi-classical limit so as to shed some light on the meaning of the specularity parameter $p$ introduced in the Fuchs-Sondheimer method. To this end, we make two conventional approximations, namely, 
\begin{equation}
\left(f(\varepsilon')-f(\varepsilon)\right)/(\varepsilon' - \varepsilon) \approx f'_0(\varepsilon), \quad \hbar\bar{\omega} + \varepsilon_{\mathbf{k}q} - \varepsilon_{\mathbf{k}+\delta\mathbf{k} q+\delta q} \approx \hbar\tilde{\omega} - \hbar \delta q v_z,
\end{equation}
where $\tilde{\omega} = \bar{\omega} - \delta \mathbf{k}\cdot \mathbf{v}_\parallel$, with $\mathbf{v}_\parallel = \hbar \mathbf{k}/m$ and $v_z = \hbar q/m$. With these approximations, Eq.~(\ref{34}) can be recast as
\begin{equation}
\sigma^{\mu\nu}_{\delta\mathbf{k}}(z,z') \approx \frac{1}{i\mathcal{V}} \sum_{\mathbf{k}q} f'_0(\varepsilon_{\mathbf{k}q}) \mathcal{L}^{\mu\nu}_{\delta \mathbf{k}}(z,z';\mathbf{k}q), \quad \mathcal{L}^{\mu\nu}_{\delta \mathbf{k}}(z,z';\mathbf{k}q) = \int^\infty_{-\infty} \frac{d \delta q}{2\pi}  \frac{j^\mu_{\mathbf{k}q,\mathbf{k}+\delta\mathbf{k} q+\delta q}(z) j^\nu_{\mathbf{k}+\delta\mathbf{k} q+\delta q,\mathbf{k}q}(z')}{\tilde{\omega} - v_z \delta q}. \label{36}
\end{equation}
where the sum in (\ref{34}) over $\delta q$ has been converted into an integration as usual. 

To make progress, let us confine ourselves to the bulk region of the metal. Inserting Eq.~(\ref{23}) in (\ref{31}), we find
\begin{eqnarray}
&~& \mathbf{j}^{\Psi}_{\mathbf{k}q,\mathbf{k}+\delta\mathbf{k} q+\delta q}(z) = e 
\begin{pmatrix}
(v_x + \delta v_x/2) \left(e^{-i\delta q z} + R^*_{\mathbf{k}q}R_{\mathbf{k}+\delta\mathbf{k} q+\delta q} e^{i\delta q z}\right)\\
(v_y + \delta v_y/2) \left(e^{-i\delta q z} + R^*_{\mathbf{k}q}R_{\mathbf{k}+\delta\mathbf{k} q+\delta q} e^{i\delta q z}\right)\\
(v_z + \delta v_z/2) \left( - e^{-i\delta q z} + R^*_{\mathbf{k}q}R_{\mathbf{k}+\delta\mathbf{k} q+\delta q} e^{i\delta q z}\right)
\end{pmatrix}
+ eR^*_{\mathbf{k}q}
\begin{pmatrix}
v_x + \delta v_x/2 \\
v_y + \delta v_y/2 \\
- \delta v_z/2
\end{pmatrix}
e^{-i(2q+\delta q)z} + eR_{\mathbf{k}+\delta\mathbf{k} q+\delta q} 
\begin{pmatrix}
v_x + \delta v_x/2 \\
v_y + \delta v_y/2 \\
\delta v_z/2
\end{pmatrix}
e^{i(2q+\delta q)z} \nonumber \\
&~& \quad \quad \quad \quad \quad \quad \quad \quad \approx 
e 
\begin{pmatrix}
v_x \left(e^{-i\delta q z} + \abs{R_{\mathbf{k}q}}^2 e^{i\delta q z}\right) \\
v_y \left(e^{-i\delta q z} + \abs{R_{\mathbf{k}q}}^2 e^{i\delta q z}\right) \\
v_z \left( - e^{-i\delta q z} + \abs{R_{\mathbf{k}q}}^2 e^{i\delta q z}\right)
\end{pmatrix}
+ eR^*_{\mathbf{k}q}
\begin{pmatrix}
v_x \\
v_y \\
-\delta v_z/2
\end{pmatrix}
e^{-i(2q+\delta q)z} + eR_{\mathbf{k} q} 
\begin{pmatrix}
v_x \\
v_y \\
\delta v_z/2
\end{pmatrix}
e^{i(2q+\delta q)z}. \label{37}
\end{eqnarray}
Here $\delta \mathbf{v} = (\hbar/m) (\delta\mathbf{k},\delta q)$ and in the second line we have assumed that $R_{\mathbf{k}+\delta\mathbf{k} q+\delta q} \approx R_{\mathbf{k}q}$. The last two terms, which are linear in the reflection amplitude, signify interference between the incident wave and the specularly reflected wave. These terms are typically ignored in the semi-classical limit, see below. In  addition, by inserting Eq.~(\ref{23}) in (\ref{33}), we find
\begin{equation}
\begin{pmatrix}
\mathcal{P}^\parallel_{\mathbf{k}q,\mathbf{k}+\delta\mathbf{k} q+\delta q}(z) \\
\mathcal{P}^z_{\mathbf{k}q,\mathbf{k}+\delta\mathbf{k} q+\delta q}(z)
\end{pmatrix}
= \sum_{\mathbf{k}_d} S^*_K(\mathbf{k},\mathbf{k}+\mathbf{k}_d)S_{K'}(\mathbf{k}+\delta\mathbf{k},\mathbf{k}+\delta\mathbf{k}+\mathbf{k}_d) e^{i \delta q_s z} 
\begin{pmatrix}
1 \\
\frac{q_s+q'_s}{q+q'} 
\end{pmatrix} 
\approx
\sum_{\mathbf{k}_d} \abs{S_K(\mathbf{k},\mathbf{k}+\mathbf{k}_d)}^2 e^{i \delta q_s z} 
\begin{pmatrix}
1 \\
\frac{q_s}{q} 
\end{pmatrix}, \label{38}
\end{equation}
where $K' = \sqrt{(\mathbf{k} + \delta \mathbf{k})^2+(q+\delta q)^2}$ and $\delta q_s = q'_s - q_s$ with $q_s = \sqrt{K^2 - (\mathbf{k}+\mathbf{k}_d)^2}$ and $q'_s = \sqrt{K^{'2} - (\mathbf{k}+\delta\mathbf{k} + \mathbf{k}_d)^2}$. The quantity $\abs{S_K(\mathbf{k},\mathbf{k}+\mathbf{k}_d)}^2 (q_s/q)$ has a simple physical content: it is the ratio of the flux through a plane parallel to the surface carried by the reflected wave with wave vector $\mathbf{k}+\mathbf{k}_d$ to that carried by the incident wave. Obviously, one must have
\begin{equation} 
\sum_{\mathbf{k}_d}\abs{S_K(\mathbf{k},\mathbf{k}+\mathbf{k}_d)}^2 (q_s/q) = 1 - ~ \abs{R_{\mathbf{k}q}}^2. \label{fr}
\end{equation}
Note that $\delta q_s = 2(q \delta q - \mathbf{k}_d\cdot\delta\mathbf{k})/(q_s+q'_s) \approx \delta q - \delta\mathbf{k}\cdot\mathbf{k}_d/q$. With this, Eq.~(\ref{38}) can be rewritten as
\begin{equation}
\begin{pmatrix}
\mathcal{P}^\parallel_{\mathbf{k}q,\mathbf{k}+\delta\mathbf{k} q+\delta q}(z) \\
\mathcal{P}^z_{\mathbf{k}q,\mathbf{k}+\delta\mathbf{k} q+\delta q}(z)
\end{pmatrix}
\approx e^{i\delta q z} 
\begin{pmatrix}
\mathcal{P}^\parallel_{\mathbf{k}q}(z) \\
\mathcal{P}^z_{\mathbf{k}q}(z)
\end{pmatrix}, \quad
\begin{pmatrix}
\mathcal{P}^\parallel_{\mathbf{k}q}(z) \\
\mathcal{P}^z_{\mathbf{k}q}(z)
\end{pmatrix}
=
\sum_{\mathbf{k}_d} \abs{S_K(\mathbf{k},\mathbf{k}+\mathbf{k}_d)}^2 e^{-i(\delta\mathbf{k}\cdot\mathbf{k}_d/q)z} 
\begin{pmatrix}
1 \\
\frac{q_s}{q} 
\end{pmatrix}. \label{p}
\end{equation}
Using Eq.~(\ref{fr}), one can infer that $\mathcal{P}^{\parallel}_{\mathbf{k}q}$ and $\mathcal{P}^{z}_{\mathbf{k}q}$, which are real if $S_K$ depends only on the magnitude of $\mathbf{k}_d$ as assumed before, are always less than $1- ~\abs{R_{\mathbf{k}q}}^2$ due to the oscillatory factors $\exp\left(-i(\delta\mathbf{k}\cdot \mathbf{k}_d)z\right)$. Substituting the expression in Eq.~(\ref{32}), we immediately get $\mathbf{j}^{d}_{\mathbf{k}q,\mathbf{k}+\delta\mathbf{k} q+\delta q}(z)$, which combines with Eq.~(\ref{37}) to yield
\begin{equation}
\mathbf{j}_{\mathbf{k}q,\mathbf{k}+\delta\mathbf{k} q+\delta q}(z)
\approx e
\begin{pmatrix}
v_x \Gamma^x_{\mathbf{k}q}(z;\delta q)\\
v_y \Gamma^y_{\mathbf{k}q}(z;\delta q)\\
v_z \Gamma^z_{\mathbf{k}q}(z;\delta q)
\end{pmatrix}
+
eR^*_{\mathbf{k}q}
\begin{pmatrix}
v_x \\
v_y \\
-\delta v_z/2
\end{pmatrix}
e^{-i(2q+\delta q)z} + eR_{\mathbf{k} q} 
\begin{pmatrix}
v_x \\
v_y \\
\delta v_z/2
\end{pmatrix}
e^{i(2q+\delta q)z}. \label{ints}
\end{equation}
where $\Gamma^x_{\mathbf{k}q}(z;\delta q) = \Gamma^y_{\mathbf{k}q}(z;\delta q) = e^{-i\delta qz} + e^{i\delta qz} \left(\abs{R_{\mathbf{k}q}}^2 + \mathcal{P}^\parallel_{\mathbf{k}q}(z)\right)$ and $\Gamma^z_{\mathbf{k}q}(z;\delta q) = -e^{-i\delta qz} + e^{i\delta qz} \left(\abs{R_{\mathbf{k}q}}^2 + \mathcal{P}^z_{\mathbf{k}q}(z)\right)$. As aforementioned, the last two terms in Eq.~(\ref{ints}), which stand for the interference between the specularly reflected wave and the incident wave, are usually ignored in the semi-classical limit. This can be achieved by performing a further coarse-graining of $\mathbf{j}_{\mathbf{k}q,\mathbf{k}+\delta\mathbf{k} q+\delta q}(z)$ over $z$. In other words, one approximates $\overline{e^{\pm i (2q+\delta q)z}} \approx 0$. To see this, it suffices to notice that $q = mv_z/\hbar$ and the integral in (\ref{36}), which can be carried out by the technique of contour integration, virtually sets $\delta q = \tilde{\omega}/v_z$ and hence $\exp\left(\pm i(2q + \delta q)z\right)$ oscillates faster than $\exp\left(\sqrt{2m\omega/\hbar}\right)$, which oscillates faster than $\exp(ik_Fz)$ for $\omega$ as small as a millionth electron volt. As such, we obtain the coarse-grained conductivity in Eq.~(\ref{36}), still to be called $\sigma^{\mu\nu}_{\delta\mathbf{k}}(z,z')$, as follows,
\begin{equation}
\sigma^{\mu\nu}_{\delta\mathbf{k}}(z,z') = e^2\left(\frac{m}{2\pi \hbar}\right)^2 \int_> d^3\mathbf{v} \left(-f'_0(\varepsilon)\right) \frac{v_\mu \Gamma^{\mu\nu}_{\delta \mathbf{k}}(\mathbf{v};z,z')v_\nu}{v_z}, \quad \Gamma^{\mu\nu}_{\delta \mathbf{k}}(\mathbf{v};z,z') = \frac{1}{2\pi i} \int^\infty_{-\infty} d \delta q ~ \frac{\Gamma^\mu_{\mathbf{k}q}(z;\delta q)\Gamma^{\nu*}_{\mathbf{k}q}(z';\delta q)}{\tilde{\omega}/v_z - \delta q},\label{41}
\end{equation}
where we have converted the sum over $\mathbf{k}q$ into an integral over $\mathbf{v}$ with $v_z\geq0$ (indicated by the symbol ``$>$"), noting that $\mathbf{v} = (\hbar/m) (\mathbf{k},q)$ and $q\geq0$, and the matrix
\begin{equation}
\Gamma^{\mu\nu}_{\delta \mathbf{k}}(\mathbf{v};z,z') = 
\begin{cases}
\alpha_\parallel\Theta(z-z') e^{-i\frac{\tilde{\omega}}{v_z}(z'-z)} + \Theta(z'-z) e^{i\frac{\tilde{\omega}}{v_z}(z'-z)} + p_\parallel e^{i\frac{\tilde{\omega}}{v_z}(z'+z)}, & \text{if} ~ ~ \mu=x,y; \nu = x,y \\
\alpha_z\Theta(z-z') e^{-i\frac{\tilde{\omega}}{v_z}(z'-z)} + \Theta(z'-z) e^{i\frac{\tilde{\omega}}{v_z}(z'-z)} - p_z e^{i\frac{\tilde{\omega}}{v_z}(z'+z)}, & \text{if} ~ ~ \mu = z; \nu=z \\
\alpha_{\parallel z}\Theta(z-z') e^{-i\frac{\tilde{\omega}}{v_z}(z'-z)} - \Theta(z'-z) e^{i\frac{\tilde{\omega}}{v_z}(z'-z)} - p_\parallel e^{i\frac{\tilde{\omega}}{v_z}(z'+z)}, & \text{if} ~ ~ \mu = x,y; \nu=z \\
\alpha_{z\parallel}\Theta(z-z') e^{-i\frac{\tilde{\omega}}{v_z}(z'-z)} - \Theta(z'-z) e^{i\frac{\tilde{\omega}}{v_z}(z'-z)} + p_z e^{i\frac{\tilde{\omega}}{v_z}(z'+z)}, & \text{if} ~ ~ \mu = z; \nu=x,y.
\end{cases}
\label{42}
\end{equation}
Here a few coefficients have been introduced, which are defined by~\cite{note6,note7}
\begin{equation}
p_\parallel = \abs{R_{\mathbf{k}q}}^2 + \overline{\mathcal{P}^\parallel_{\mathbf{k}q}}, \quad p_z = \abs{R_{\mathbf{k}q}}^2 + \overline{\mathcal{P}^z_{\mathbf{k}q}}, \quad \alpha_\parallel  = p^2_{\parallel}, \quad \alpha_z = p^2_z, \quad \alpha_{\parallel z} = \alpha_{z \parallel} = \sqrt{\alpha_\parallel \alpha_z} = p_z p_{\parallel}. \label{43}
\end{equation} 
\end{widetext}
These quantities generally depend on $\mathbf{k}q$. 

Equations (\ref{41}) - (\ref{43}) constitute the semi-classical limit of the quantal conductivity for the bulk of a semi-infinite metal bounded by a rough surface. They are structurally the same as the results attained by the Fuchs-Sondheimer method, Eqs.~(\ref{6}) and (\ref{7}). One sees that $p_z$ and $p_\parallel$ may be defined as the quantum-mechanical version of the specularity parameter. They play a similar role here as the Fuchs parameter $p$ in the semi-classical method. Nevertheless, there are two basic differences. Firstly, two parameters -- instead of one -- are needed in the quantum theory, one of them ($p_z$) characterizing the electrical responses to a field normal to the surface while the other ($p_\parallel$) to field along the surface. This reflects on the intrinsic anisotropy associated with the presence of a surface, which is not captured in the semi-classical theory. Secondly, the parameters $p_z$ and $p_\parallel$ do not generally measure the fraction of specularly reflected electrons: they include contributions from both specularly and diffusively reflected electrons, in contrast to the interpretation of Fuchs and others. However, the contributions from diffusively reflected electrons are strongly suppressed by the oscillatory factor $\exp\left(-i(\delta\mathbf{k}\cdot\mathbf{k}_d/q)z\right)$ [c.f. Eq.~(\ref{p}) and Ref.~\cite{note7}], which vanishes upon coarse-graining if $\Lambda \delta\mathbf{k}\cdot\mathbf{k}_d/q >1$. It is clear that the larger $\Lambda$ is, the stronger the suppression shall become. In the study of anomalous skin effect and other phenomena that uses a uniform electric field, one can take $\Lambda$ to be very large and then the contribution from diffusively reflected electrons can be reduced to a negligible level. In such cases, both $p_z$ and $p_\parallel$ become equal to $\abs{R}^2$ and thus indeed give the fraction of specularly reflected electrons. 

Finally, it should be seen that the reasoning that brings about Eq.~(\ref{16}) applies here as well. 

\section{Discussions and summary}
\label{sec:4}
Clarifying surface effects on electronic conduction in metals is of persistent interest, as such effects inevitably affect a plethora of fundamental physical processes underlying common experimental techniques employed for probing materials, examples including light scattering (e.g. ellipsometry) and particle scattering (e.g. energy loss spectroscopy) techniques. An adequate understanding of these effects constitutes a prerequisite for reliably interpreting empirical observations and extracting material properties. 

The semi-classical Fuchs-Sondheimer theory has been the first and remains the most convenient approach to enlightening experimental data involving surface effects. This is largely due to its simplicity: the theory builds in apparent translation symmetry -- even for rough surfaces -- in its basic equation and hence requires a single parameter $p$, which supposedly possesses an intuitive physical meaning, to capture surface scattering effects. Nevertheless, despite the long history on this subject dated back to the 1930s and extensive use of this method in a wide context, a quantum mechanical version of this theory has hitherto not been derived. The present work fills this gap of understanding. Starting from the quantal Kubo-Greenwood formalism, we show that the conductivity of a semi-infinite metal can be reduced to the same structure as obtained in the Fuchs-Sondheimer theory, the main difference being that, instead of a single parameter $p$, two parameters, $p_\parallel$ and $p_z$, are required to capture the scattering effects [c.f. Eqs.~(\ref{7}) and (\ref{42})]. These parameters are further expressed in terms of the fundamental quantum mechanical scattering matrix [c.f. Eq.~(\ref{scaM})] for an arbitrary surface, and thus their physical meanings become clear. It turns out that, $p_\parallel$ and $p_z$ are contributed from both specularly reflected electrons and diffusely reflected one. Indeed, the discrepancy between them is entirely due to the diffusely reflected electrons. However, contribution from the diffusely reflected electrons is strongly suppressed on a scale much larger than the lattice constant $a$, as is so in for example optical and transport experiments. In such case, $p_\parallel$ and $p_z$ approach $p$. By fitting $p_\parallel$ and $p_z$ with empirical data, one may experimentally gauge the importance of the diffusely reflected electrons. 

Our results also vindicate the apparent translation symmetry preserved in the Fuchs-Sondheimer method, which is valid as long as the surface is macroscopically flat, that is, the roughness is perceived only on the scale of $a$. With macroscopic flatness, the interference between diffusely reflected electrons and incident as well as specularly reflected electrons vanishes on a scale larger than $a$, thence restoring the symmetry. This feature has also been confirmed by a generalized semi-classical method.

The most important message conveyed by the present work, however, is the realization that the distribution function obtained in the Fuchs-Sondheimer theory applies only to the bulk of a metal, not in the surface region. It particular, it should not be used for calculating the charge density in the surface region. This observation calls into question many existing work~\cite{flores1979,harris1972,garcia1977,katsnelson2018} on dynamical charge phenomena hosted on the surface, including surface plasma waves (SPWs) as a notable example. As far as we are concerned, virtually every work, dated back to the 1970s, employing the Fuchs-Sondheimer method for studying SPWs has wrongly assumed that the charge density in the entire metal can be calculated using the Fuchs-Sondheimer distribution function [c.f. Eqs.~(\ref{4}) and (\ref{5})]. Our recent work~\cite{deng2019,deng2017a,deng2017b,deng2017c} demonstrated that, by carefully warding off the historical fallacy, the properties of these waves can be critically modified by surface scattering. One interesting consequence is that, in the collision-less limit, SPWs may become unstable as long as the surface does not possess perfect translation symmetry.


\begin{thebibliography}{99}

\bibitem{abrikosov}
A. A. Abrikosov, \textit{Fundamentals of the Theory of Metals}, chapter 8 (Amsterdam: Elsevier, 1988).

\bibitem{ziman}
J. M. Ziman, \textit{Electrons and Phonons: The Theory of Transport Phenomena in Solids}, chapter XI (Oxford: Oxford University Press, 2001).

\bibitem{fuchs1938}
K. Fuchs, \textit{Proc. Camb. Phil. Soc.} \textbf{34}, 100 (1938).

\bibitem{sondheimer1948}
G. E. H. Reuter and E. H. Sondheimer, \textit{Proc. Roy. Soc. A} \textbf{195}, 338 (1948).

\bibitem{sondheimer1952}
E. H. Sondheimer, \textit{Adv. Phys.} \textbf{1}, 1 (1952).

\bibitem{kaganov1997}
M. I. Kaganov, G. Y. Lyubarskiy and A. G. Mitina, \textit{Phys. Rep.} \textbf{288}, 291 (1997).

\bibitem{kuan2004}
S. M. Rossnagel and T. S. Kuan, \textit{J. Vac. Sci. Technol. B} \textbf{22}, 240 (2004).

\bibitem{gall2007}
J. M. Purswani and D. Gall, \textit{Thin Solid Films} \textbf{516}, 465 (2007).

\bibitem{gall2009}
J. S. Chawla and D. Gall, \textit{Appl. Phys. Lett.} \textbf{94}, 252101 (2009).

\bibitem{gall2008}
V. Timoshevskii, Y. Ke, H. Guo and D. Gall, \textit{J. Appl. Phys.} \textbf{103}, 113705 (2008).

\bibitem{ke2009}
Y. Ke, F. Zahid, V. Timoshevskii, K. Xia, D. Gall and H. Guo, \textit{Phys. Rev. B} \textbf{79}, 155406 (2009).

\bibitem{camley1989}
P. E. Camley and J. Barnas, \textit{Phys. Rev. Lett.} \textbf{63}, 664 (1989).

\bibitem{barnas1990}
J. Barnas, A. Fuss, R. E. Camley, P. Gr\"{u}nberg and W. Zinn, \textit{Phys. Rev. B} \textbf{42}, 8110 (1990).

\bibitem{levy1990}
P. M. Levy, S. Zhang and A. Fert, \textit{Phys. Rev. Lett.} \textbf{65}, 1643 (1990).

\bibitem{flores1979}
F. Flores and F. Garcia-Moliner, \textit{J. Phys. C.: Solid State Phys.} \textbf{12}, 907 (1979).

\bibitem{harris1972}
J. Harris, \textit{J. Phys. C: Solid State Phys.} \textbf{5}, 1757 (1972).

\bibitem{garcia1977}
F. Garcia-Moliner and F. Flores, \textit{J. Phys.} \textbf{38}, 851 (1977).

\bibitem{deng2019}
H.-Y. Deng, \textit{New. J. Phys.} \textbf{21}, 043055 (2019).

\bibitem{deng2017a}
H.-Y. Deng, K. Wakabayashi and C.-H. Lam, \textit{Phys. Rev. B} \textbf{95}, 045428 (2017).

\bibitem{deng2017b}
H.-Y. Deng, \textit{Phys. Rev. B} \textbf{95}, 125442 (2017).

\bibitem{deng2017c}
H.-Y. Deng, \textit{J. Phys. Condens. Matter} \textbf{29}, 455002 (2017).

\bibitem{deng2018}
H.-Y. Deng, \textit{ArXiv}: 1806.08308 (2018). 

\bibitem{katsnelson2018}
A. Principi, E. van Loon, M. Polini and M. I. Katsnelson, \textit{Phys. Rev. B} \textbf{98}, 035427 (2018). 

\bibitem{kubo}
R. Kubo, \textit{J. Phys. Soc. Jpn.} \textbf{12}, 570 (1957).

\bibitem{greenwood}
D. A. Greenwood, \textit{Proc. Phys. Soc. London} \textbf{71}, 585 (1958).

\bibitem{zhang1992}
S. Zhang, P. M. Levy and A. Fert, \textit{Phys. Rev. B} \textbf{45}, 8689 (1992).

\bibitem{tesanovic1986}
Z. Tesanovic, M. Jaric and S. Maekawa, \textit{Phys. Rev. Lett.} \textbf{57}, 2760 (1986).

\bibitem{chambers1950}
R. G. Chambers, \textit{Proc. R. Soc. London Ser. A} \textbf{202}, 378 (1950).

\bibitem{chambers1952}
R. G. Chambers, \textit{Proc. R. Soc. A} \textbf{65}, 458 (1952).

\bibitem{camblong1992}
H. E. Camblong and P. M. Levy, \textit{Phys. Rev. Lett.} \textbf{69}, 2835 (1992).

\bibitem{camblong1993}
H. E. Camblong and P. M. Levy, \textit{J. Appl. Phys.} \textbf{73}, 5533 (1993).

\bibitem{camblong1995}
H. E. Camblong, \textit{Phys. Rev. B} \textbf{51}, 1855 (1995).

\bibitem{zhang1995}
X.-G. Zhang and W. H. Butler, \textit{Phys. Rev. B} \textbf{51}, 10085 (1995).

\bibitem{note1}
Here the electric field is the sum of an external electric field and the mean field engendered by the electrons themselves. The conductivity to be calculated is thus the bare conductivity excluding many-body effects.

\bibitem{note}
In establishing this equation, we have used the fact that $\left(\mathbf{v}\cdot\partial_\mathbf{x} + \frac{\mathbf{F}_s}{m}\cdot \partial_\mathbf{v}\right) f_0 = 0$.

\bibitem{note2}
By coarse-graining of a function $O(x)$, we mean an average of $O(x)$ over a region $\Omega_x$ centered about $x$ with a linear dimension ($\sim \Lambda$) much larger than the length scale for diffusive scattering ($\sim a$), i.e. $\overline{O(x)} = \int_{\Omega_x} dx' O(x') / \int_{\Omega_x} dx'$. This removes any Fourier components of $O$ that vary significantly within that scale.  

\bibitem{note3}
To be more accurate, we may insert here a factor $\alpha \sim 1$ in front of $f'_0$, defined by $$\int^z_{-z_s/2} dz' ~ e^{i\frac{\tilde{\omega}(z-z')}{v_z}} \mathbf{E}_{\mathbf{k}}(z') f'_{0,0}(z',\mathbf{v}) \approx \alpha f'_{0} \int^z_{-z_s/2} dz' ~ e^{i\frac{\tilde{\omega}(z-z')}{v_z}} \mathbf{E}_{\mathbf{k}}(z').$$ Here $z>z_s/2$ is understood.

\bibitem{note4}
One might think that $\rho$ calculated with $g$ can analogously be related to $\tilde{\rho}$ calculated with $g^{(b)}$ via a phenomenological function, say $w(z)$. However, on the scale of $\Lambda\gg z_s$, $w(z)$ may not degenerate with $\Theta(z)$. Rather, $w(z)$ can be singular like a Dirac delta function. For example, if there are some charges in the surface layer, the density of these charges will appear like a delta function in the limit where the layer appears infinitely thin.

\bibitem{notekubo}
In Eq.~(\ref{24}), $\bar{\omega} = \omega + i/\tau$. In a clean metal free from impurities and other type of electronic collisions, one may take the limit of $1/\tau$ being a positive infinitesimal. In general, we view $1/\tau$ as the imaginary part of the electron self-energy $\Sigma$ arising from the collisions, see for instance Ref.~\cite{zhang1995}. 

\bibitem{note6}
We should note that these coefficients are functions of $z$ and $z'$. Namely, $p_\parallel(z) = \abs{R_{\mathbf{k}q}}^2 + \overline{\mathcal{P}^\parallel_{\mathbf{k}q}(z)}$ and $p_z(z) = \abs{R_{\mathbf{k}q}}^2 + \overline{\mathcal{P}^z_{\mathbf{k}q}(z)}$, as well as $\alpha_\parallel(z,z') = p_\parallel(z)p_\parallel(z')$, $\alpha_z(z,z') = p_z(z)p_z(z')$, $\alpha_{\parallel z}(z,z') = p_\parallel(z)p_z(z')$ and $\alpha_{z\parallel}(z,z') = p_z(z)p_\parallel(z')$. However, the dependence should be weak if $\Lambda$ is large. 

\bibitem{note7}
It is useful to see that $p_\parallel$ and $p_z$ can also be written as
$$\left(p_\parallel,p_z\right) = \sum_{\mathbf{k}_s} \abs{S_K(\mathbf{k},\mathbf{k}_s)}^2 \overline{e^{-i(\delta\mathbf{k}\cdot\mathbf{k}_s/q)z}}\left(1,q_s/q\right).$$ Here the sum over $\mathbf{k}_s$ includes specularly and diffusively reflected electrons. For the former, $\mathbf{k}_s = 0$ and $q_s/q=1$. This expression puts them on equal footing. The contributions from diffusively reflected electrons are suppressed by the oscillatory factor, which is one for specularly reflected electrons. 

\end{thebibliography}
\end{document}